\documentclass{article}
\usepackage{amssymb}
\usepackage{amsmath}

\newtheorem{theorem}{Theorem}

\newtheorem{lemma}[theorem]{Lemma}

\newenvironment{proof}[1][Proof]{\noindent\textbf{#1.} }{\ \rule{0.5em}{0.5em}}
\input{tcilatex}

\begin{document}

\begin{center}
{\Huge What is Probability?}

\bigskip

{\LARGE Simon Saunders}

\medskip

\textit{Philosophy of Physics Group, University of Oxford}
\end{center}

What is probability? Physicists, mathematicians, and philosophers have been
engaged with this question since well before the rise of modern physics. But
in quantum mechanics, where probabilities are associated only with
measurements, the question strikes to the heart of other foundational
problems: what distinguishes measurements from other physical processes? Or
in more formal terms: when are the unitary dynamical equations suspended in
favour of probabilistic ones?

This is the \textit{problem of measurement} in quantum mechanics. The most
clear-cut solutions to it change the theory: they either add hidden
variables (as in the pilot-wave theory), or they give up the unitary
formalism altogether (as in state-reduction theories). The two strategies
are tied to different conceptions of probability: probability as in
classical statistical mechanics (as formulated by Boltzmann, Gibbs and
Einstein), and probability as in Brownian motion (with the dynamics given by
a stochastic process, as formulated by Einstein and Smulochowski). The
former is sometimes called \textit{epistemic}\ probability, as classical
mechanics is deterministic: probabilities arise as a consequence of
incomplete description. It is the latter, stochastic, probability that is
usually thought the more fundamental, as it enters directly into the
fundamental equations.

Indeterminism encouraged a late-comer to the philosophy of probability,
Popper's \textit{propensity}\ approach \cite{Popper}, in which probabilities
are identified with certain kinds of properties - \textquotedblleft
dispositions\textquotedblright\ - of chance set-ups. But whether thought of
in terms of incomplete descriptions, or as propensities, these kinds of
probability have puzzled philosophers. By way of contrast, probability as 
\textit{degree of subjective expectation}, or \textit{subjective weighting},
is much clearer: if that is what probability is one can explain why it obeys
the rules that it does (I shall come back to this point later). Neither the
epistemic probabilities of classical statistical mechanics, nor the
objective probabilities of a stochastic dynamics as in Brownian motion, are
happily thought of as subjective expectations. They are grounded, surely, in
facts about physical states of affairs, independent of persons altogether.
It is true that epistemic probability is often thought of in terms of
ignorance, or incomplete knowledge, but it is precisely this link with
purely subjective considerations that is so puzzling from the point of view
of its role in thermodynamics: heat transfers, surely, take place according
to thermodynamical laws, independent of whether anyone is looking, and of
what anyone knows. The point is all the more evident in the case of
stochastic probability. These epistemic and stochastic probabilities of
physics are objective rather than subjective quantities. What, then, are the
physical facts that make true some probability statements but not others?
What are we trying to get right about when we make judgements of objective
probabilities? It is not informative to say it is that there are \textit{%
chances}\ that are thus and so; the difficulty, that has long bedevilled the
philosophy of probability, is to say what chances could possibly be.

There are well-known, failed, candidates for the role. The most commonplace
looks to the \textit{evidence} for probability statements, the observed
statistical data. Of course if the question is what chances are, rather than
what we believe them to be, we should consider unobserved data as well; in
the most simple-minded approach, these two are simply identified -
specifically, chances are identified with \textit{long-run} relative
frequencies. But the objections to doing this are obvious. How long is long
enough? In the short-run we cannot expect chance and relative frequencies to
line up exactly, whilst the infinite limit is never actually reached; so
what, precisely, are the chances? Presumably a function of the number of
trials and the actual relative frequencies; but what function?

One can prove that for repeated trials the relative frequencies of outcomes
will approximately match the probability of each outcome (assumed
independent), but only in the sense that if $p(n,\delta )$ is the
probability that they differ by more than $\delta >0,$ for $n$ trials, then $%
\underset{n\rightarrow \infty }{\lim }p(n,\delta )=0$ (the \textit{law of
large numbers}). But this is not to identify probabilities with any actual
relative frequencies; so long as $n$ is finite, we still have to do with
probabilities (the quantities $p(n,\delta )$). What this and the various
other laws of this kind show, rather, is that the concept of probability
enters into the very relation between statements of evidence and statements
about chance; statistical evidence bears out a probability claim only to
some degree - where \textquotedblleft some degree\textquotedblright\ is some
degree of probability.

Could the latter probability be subjective rather than objective
probability? It could - but only if we assume that subjective probability,
at least in such cases, agrees with chance: for we calculate the probability
for matching statistics with chances by using those same chances. What
underwrites this assumption? This is a question that arises even if one knew
what chances really are; why should subjective expectations be set equal to
objective probabilities? Put to one side worries about long-run frequencies
not being long enough; why, given that the long-run frequency of some
outcome is thus and so, should we subjectively expect that outcome to that
degree on the very next trial?

We have two questions:

\begin{enumerate}
\item What, physically, is objective probability (chance)?

\item Why should subjective probability track chance?
\end{enumerate}

\noindent Because of its importance in the philosophy of probability and of
rational action, the principle that subjective probability should track
chance - or the slightly more involved principle that it should, however
much additional information one may have, short of knowledge of what the
actual outcome will be, has been dubbed the \textit{principal principle} 
\cite{Lewis}. We are sure this principle is true, but we are at a loss to
say why. Indeed, failing an account of what chances are, it is hard to see
how the principle \textit{could} be justified; for it ought to be facts
about physical states of affairs that dictate our subjective expectations of
future contingencies. What are those facts? The two questions are
interdependent.

One can rest on the authority of science. One can say that it is a
requirement of any theory of rationality that our beliefs should be based on
our best scientific theories. One can take it as an extension of this that
we should tailor our subjective expectations of chance outcomes to whatever
our best theory says are their chances; the principal principle would then
be part of any theory of rational action worthy of the name. Maybe so, but
then it will be an unexplained part. It is also a very large part of the
theory of practical reasoning, perhaps the largest part (the part that deals
with physical contingencies). One would, in the light of our best theories
in physics, like to do better. The challenge is to say what it is about the
world that makes statements of objective probability true, and why, given
such states of affairs, we should act accordingly, with subjective
probabilities fixed by the objective ones.

Questions (1) and (2) are now the most important ones in the philosophy of
probability. It was not always so; Popper, when he proposed to abandon the
link between long-run frequency and chance, wanted an account of probability
that made sense of the single-case and that made no reference to human
knowledge. Worthy aims; but to suppose that the chance of an outcome of an
experiment is a \textquotedblleft disposition\textquotedblright\ of that
experiment (and more generally, that the chance of an event given a certain
chance set-up is a \textquotedblleft disposition\textquotedblright\ of that
set-up) in itself solves very little; for what, physically, are these
\textquotedblleft dispositions\textquotedblright ?\ Popper was never able to
say in terms of categorical physical properties - properties that are not
themselves \textquotedblleft chancy\textquotedblright\ or equally in need of
explanation. Nor was he able to in classical physics. Even in classical
games of chance, where chances are directly related to the symmetries of a
chance set-up - the six faces of a die, the two sides of a coin - dynamics
comes into it. Throw the die or the coin just so, and the statistics, the
evidence for the chances, can be anything you like. \ The dynamics, it
seems, can override chance.

This worry would seem to arise in any deterministic theory. It is made
marginally more palatable by putting it in terms of initial conditions
rather than dynamics (that certain initial conditions are \textquotedblleft
typical\textquotedblright ). Better still, adopt the picture of a
probability distribution as a measure on an ensemble (usually infinite) of
hypothetical physical systems, a picture in which dynamics is absent
altogether. One can understand its appeal, but only from this narrow
perspective. What it is about the actual world that makes the probabilities
what they are is never explained. Why we are supposed to have expectations
about the real world because of the values of a measure on a fictitious
ensemble of imaginary worlds is never explained.

(1) is evidently a hard question, but it is at least a physical question; it
is not so clear that (2) is. It would be understandable were physicists to
limit themselves to (1), to be answered by physics as usual (as in saying
what temperature really is, or what solidity really is); (2) can be left to
the philosophers. But we are interested in probability, above all, in
quantum mechanics, and there one is up against the problem of measurement.
When it comes to the problem of measurement physics is not its usual self.
As we shall see, it turns out that the best answer to (1) so far in evidence
also provides an answer to (2) - but at a price.

Our story must proceed in stages. We begin with orthodoxy; next we consider
alternatives to quantum mechanics. From then on we consider probability from
the point of view of decoherence theory and the unadulterated formalism.

\section{Orthodoxy}

\subsection{Gleason's Theorem}

Consider a system $a$ with Hilbert space $H^{a}$ and inner product $<.,.>.$
For $\phi \in H^{a}$ and any projector $\widehat{P}$ on $H^{a}$, the
probability $\mu (\phi ,\widehat{P})$ is defined by the rule (\textit{the
Born rule}): 
\begin{equation*}
\widehat{P}\text{ is measured in }\phi \text{ }\Rightarrow \text{ }\mu (\phi
,\widehat{P})=\frac{<\phi ,\widehat{P}\phi >}{<\phi ,\phi >}.
\end{equation*}%
From the RHS we see that $\mu $ is additive and indeed countably additive
over any partitioning $\{\widehat{P}_{k}\}$ of $H^{a}$ (any pair-wise
orthogonal set of projectors summing to the identity). Looking to the LHS,
we suppose that each partitioning corresponds to a particular kind of
experiment (one that \textit{measures} $\widehat{P}$). This is where the
\textquotedblleft intent\textquotedblright\ of the experiment or the notion
of the \textquotedblleft observation\textquotedblright\ that is made comes
in; it has proved to be very hard to make do with a purely physical
specification of the apparatus instead.

We need a few more definitions in order to state Gleason's theorem. Given a
density matrix $\rho $ (positive, self-adjoint, and of trace one) in place
of a vector in $H^{a},$ the rule is:

\begin{equation*}
\widehat{P}\text{ is measured in }\rho \text{ }\Rightarrow \text{ }\mu (\rho
,\widehat{P})=Tr(\rho \widehat{P})
\end{equation*}

\noindent where $Tr$ is the trace, yielding a weighted sum over
probabilities as defined for the pure case. The set of all projectors on $H$
has an algebraic structure defined by subspace inclusion (a partial
ordering). Using it one can define the meet (\textquotedblleft
and\textquotedblright ) and join (\textquotedblright or\textquotedblright )
operations, and under these it is a lattice. It is not of course a Boolean
lattice (for which the meet and join operations are distributive), but it
has (infinitely) many Boolean sublattices, and it is natural \ to demand
that a probability function on this lattice should be additive on each
Boolean sublattice of the total lattice. This is the condition of Gleason's
celebrated theorem \cite{Gleason}:

\begin{theorem}[Gleason]
Let $f$ be any function on projectors on a Hilbert space $H$ of dimension $%
d>2$ to the unit interval which is additive for any set of pairwise disjoint
projectors on $H.$ Then there exists a unique density matrix $\rho $ such
that for any $\widehat{P}$ on $H,$ $f(\widehat{P})=Tr(\rho \widehat{P}).$
\end{theorem}

Gleason's theorem is a derivation of part of the Born rule, but of course it
says nothing about \textquotedblleft measurements\textquotedblright\ or
\textquotedblleft experiments\textquotedblright ; nor, on reflection, is the
premise of the theorem so clearly motivated. It is assumed that the
probability for an outcome $\widehat{P}$ belonging to one sublattice is the
same when $\widehat{P}$ is considered as a member of another. \ The
assumption appears innocuous, but it has non-trivial consequences. For
example, let $\widehat{P}_{i}$ project onto the subspace spanned by $\chi
_{i}\in H^{a}$, $i=1,2$, and let $\widehat{P}_{\pm }$ project onto $\chi
_{1}\pm \chi _{2}$; then $\{\widehat{P}_{1},\widehat{P}_{2}\}$ generates one
Boolean sublattice and $\{\widehat{P}_{+},\widehat{P}_{-}\}$ another. Yet if 
$f(\widehat{P}_{1})=f(\widehat{P}_{2})=0$, $\ $then by additivity $f(%
\widehat{P}_{1}+\widehat{P}_{2})=f(\widehat{P}_{+}+\widehat{P}_{-})=0$; and
since $f$ is positive, from additivity again it follows that $f(\widehat{P}%
+)=f(\widehat{P}_{-})=0.$ So probabilities for one family of projectors
constrain those for another, even though the two do not commute ($[\widehat{P%
}_{i},\widehat{P}\pm ]\neq 0$)$.$ Should constraints like this be imposed
that relate measurements on non-commuting operators? Like that other
celebrated theorem in the foundations of quantum mechanics (von Neumann's
no-go\ theorem for hidden variables) the condition of Gleason's theorem may
be physically unmotivated.

Of course the premise can be taken as the expression of a phenomenological
principal. It is true as goes the statistics of actual experiments; the
statistics of an outcome are the same whatever other quantity (so long as it
is represented by a commuting operator) is measured. Gleason's theorem, we
may suppose, shows us how a phenomenological principle implies a certain
mathematical representation of probabilities in quantum mechanics, rather as
Kelvin showed how thermodynamical laws imply a certain representation of
temperature; and as (although the example is a bit of a stretch) Einstein
showed how the relativity principle and the light speed principle imply a
certain representation of geometry.

The comparison with thermodynamics is not an altogether happy one. It
reminds us that statistical data may well be consistent with deterministic
theories. This point was important to Bell \cite{Bell}, who was sympathetic
to the idea of introducing hidden variables in quantum mechanics. He sought
to understand the observed statistics in terms of averages over states ruled
out by Gleason's theorem, \textit{dispersion-free} states, in which every
projector has value zero or one. And Bell noticed that the latter are
excluded by a much simpler argument than Gleason's: from the result just
proved, and from the fact that (again from additivity) if $f(\widehat{P}%
_{\chi })=1$ and if $\phi $ is orthogonal to $\chi $ then $f(\widehat{P}%
_{\phi })=0,$ it follows immediately that if $f(\widehat{P}_{\chi })=1$ and $%
f(\widehat{P}_{\phi })=0$ then $\chi $ and $\phi $ cannot be too close ($%
|\chi -\phi |>\frac{1}{2}|\phi |$). So there can be no dispersion-free
states, for if dispersion-free $f(\widehat{P}_{\chi })$ must change from $1$
to $0$ as $\chi $ is continuously rotated into $\phi $, so it must change
for vectors that are arbitrarily close.

According to Bell what is wrong with Gleason's additivity assumption, at the
level of the single case, is that it ignores a clear possibility that cannot
in principle be ruled out on experimental grounds. In the single case, the
values assigned to $\widehat{P}_{1}$ and $\widehat{P}_{2}$ may be zero (and
that can be discovered experimentally), but one cannot simultaneously
measure $\widehat{P}_{+}$ or $\widehat{P}_{-}$ so one can draw no conclusion
as to the value assigned to them in that context.

\subsection{Bohr's Copenhagen Interpretation}

Bell's reasoning was faithful to Bohr's \textit{principle of complementarity}%
, according to which quantum mechanical phenomena cannot be defined
independent of an experimental context. Given this and the fact that
experiments that measure non-commuting quantities cannot simultaneously be
performed, the way is open for results of experiments to defy classical
reasoning altogether; it is possible (this Bohr's principle of
complementarity) that results obtained by incompatible experiments cannot be
consistently fitted together according to the classical ideal of
explanation. It is this that makes room for genuine novelty, in quantum
experiments, according to Bohr. This argument was made repeatedly in Bohr's
published writings.

It is therefore embarrassing, to Bohr if not to Bell - because Bohr was out
to interpret quantum mechanics rather than to change it - that by this
reasoning a loophole is opened up in Gleason's theorem. It may be that every
dynamical quantity has a well-defined value (and every projector has the
value $0$ or $1$) but that values for commuting quantities not in fact
measured differ from the values that they would have had if they were. Such
a theory is called a \textit{contextual} hidden variable theory.

The same applies to the Kochen-Specker theorem (in effect a strengthening of
Bell's theorem). Bohr's complementarity opens a way out for dispersion-free
states in that case as well - and a way to understand the quantum
probabilities as describing only the statistics.

Leaving so much open, the Copenhagen interpretation offers no definite
account of probability and no justification for the Born rule. And even if
it could be used to underwrite Gleason's premise, it would give no answer to
(1). For according to Bohr, the state is not something physically real; the
squared modulus of the amplitude is not a categorical physical quantity. For
all that it suggests a link between (1) and (2), it does so at the expense
of (1). The Born rule gives the probabilities for the commuting family of $%
\widehat{P}$'s that the experiment is \textquotedblleft
intended\textquotedblright\ to measure. In answering (1), one wants to
dispense with the \textit{intentions} (for the probabilities we are after
are supposed to be objective rather than subjective). What is it from a
purely physical point of view about an experimental apparatus that dictates
that it is one set of probabilities that is relevant to the outcomes rather
than another? What is the correct choice of sublattice, or equivalently, of
basis?

Insisting as he did that the apparatus must be described in classical terms,
Bohr was not entirely at a loss to answer this question. Example by example,
he tried to show that the basis was dictated by some concrete feature of the
apparatus - for example, by whether or not some shutter, screen or diaphragm
was bolted to the laboratory bench. This was supposed to work in tandem with
Bohr's further thesis that in quantum mechanics one never really went beyond
classical concepts (or one or another of a complementary set of classical
concepts); that there were, in effect, \textit{no} genuinely quantum
mechanical concepts, or none that could function in explanations in the way
that classical concepts did. But here Bohr was obviously at a disadvantage;
it was part and parcel of complementarity that one could not recover the
classical ideal of explanation when it came to atomic phenomena; it was an 
\textit{a priori} prejudice on Bohr's part that genuinely quantum concepts
could never be found that would do better than the fragments of classical
physics that Bohr did vouchsafe to us.

Few were prepared to follow Bohr with his analysis of quantum mechanics, and
eventually of quantum field theory, in terms of fragments of classical
physics. Even restricted to the analysis of measurements, his systems of
trapdoors, levers and springs seemed baroque; he never was able to establish
any hard-and-fast connection between what was bolted to what and the
observable that was supposed to be measured. The doctrine of incomplete
explanation was unsuccessful in the other areas where Bohr hoped it would
deliver, in biology and psychology. It never offered any insight into the
nature of probability.

\section{Alternatives to quantum mechanics}

\subsection{Pilot-wave theory}

The alternatives to quantum mechanics are well known. To begin with
pilot-wave theory, which retains the unitary equation for the state and
supposes indeed that quantum mechanics is universal (so there is a
wave-function for the universe), one has underlying dispersion-free states,
as Bell wanted, but only for certain dynamical variables (namely
configuration space variables, the relative positions and relative
velocities). Correspondingly, one has an additional equation - the guidance
equation - that dictates the allowable trajectories through each point of
configuration space. Consistent with the Bell-Kochen-Specker theorem,
definite values are not attributed to every self-adjoint operator in quantum
mechanics, independent of context. In fact most operators aren't assigned
values at all (for example, only in certain\ contexts is any component of
spin assigned a value). And where a component of spin is assigned a value
(in the context of, say, a Stern-Gerlach experiment, for a particular
orientation of the magnetic field) no value for any other component of spin
is defined.

This is to rehabilitate Bohr's reasoning about experiments (although it can
hardly be said to lend support to his general philosophy). But the more
clear-cut interpretation of the pilot-wave theory is to suppose that the
only real physical quantities are configuration space variables (relative
positions and velocities); that all the rest are artifacts of experiments.
Dispersion-free states for these quantities are not contextual, except in
the sense that they may of course (by the non-locality of the theory) change
when the macroscopic apparatus is changed. Bell himself seemed to advocate
this position \cite{Bell2}.

So how does probability get into the picture? Much as it does in classical
statistical mechanics: one probability distribution on configuration space
is favoured (as given by the Born rule) for much the same reason that
Liouville measure on phase space is favoured in classical Hamiltonian
mechanics. The Born rule is said to be the \textquotedblleft
equilibrium\textquotedblright\ distribution (\textquotedblleft quantum
equilibrium\textquotedblright ). Once in equilibrium, systems cannot be
reliably prepared in the dispersion-free states allowed by the theory (it is
this that hides the non-locality). This smacks of conspiracy; but given
equilibrium, the situation is in one respect better than in classical
statistical mechanics. At least we are in a position to answer (1): chances
are determined by certain categorical properties in the world (the squared
norms of the components of the wave-function with respect to the position
basis) - assuming, as proponents of the pilot-wave theory usually do, that
the wave-function is physically real.

Now note the two disadvantages of this approach. The first is that if this
is the answer to Question (1), Question (2) seems entirely impenetrable. It
is hard to see why our subjective expectations should be concerned with
these amplitudes squared. The trajectory in phase space can after all be
chosen to give you any statistics you like (leaving the amplitudes
completely unchanged); if it is the trajectory we are concerned with - this
is what picks out all the actual things that happen as opposed to those that
don't - why should something completely \textit{independent }of the
trajectory, the amplitudes, be of any relevance?

The second turns this objection around. Why, in any case, suppose the
physical probabilities are given by the Born rule? Maybe they float free of
it, as classical probabilities can float free from equilibrium distributions
in statistical mechanics \cite{Valentini}. There, non-equilibrium is not
ruled out by \textit{fiat} as somehow illegitimate (classically the universe
is far from equilibrium). But if so, and there is no intrinsic connection
between physical probabilities and the amplitudes, what \textit{do}
probabilities correspond to? - and we are back to square one.

\subsection{State-reduction theory}

In pilot-wave theory state reduction is \textquotedblleft
effective\textquotedblright , as components of the state (\textquotedblleft
empty waves\textquotedblright ) irrelevant to the guidance equation can be
discarded. The alternative is to build it into the dynamics directly. Here,
unlike the case of pilot-wave theory, a considerable body of new work in
physics and mathematics was needed. The mathematical foundations were laid
by Smulochowski, Wiener and von Neumann in the 1920s. Investigations in
quantum optics in the late 1950s made use of probability measures in quantum
mechanics generalized to sequential sample spaces. They also generalized to
continuous sampling; in the mid 1960's Nelson showed how the Schr\"{o}dinger
equation may be related to a continuous Markovian stochastic process. By the
end of the 1970s, with investigations in open quantum systems theory by
Davis and his collaborators, it was clear how to construct stochastic models
to mimic the behaviour of subsystems subject to arbitrary quantum
evolutions. The first stochastic theory demonstrably equivalent to standard
quantum mechanics across a wide range of applications (all of them, however,
non-relativistic) was proposed in 1986 by Ghirardi, Rimini, and Weber. These
theories are therefore of comparatively recent origins; what do they say
about probability?

The Schr\"{o}dinger equation is replaced by a stochastic vector-valued
dynamical process. In what is perhaps the most elegant example, the \
continuous state-reduction model \cite{Ghirardi2}, this process is assumed
to be Markovian and to take the form:

\begin{equation}
d\psi =\left( \widehat{Q}(\psi )dt+\widehat{\mathbf{R}}(\psi )\cdot \mathbf{%
dB}\right) \psi .
\end{equation}

\noindent Here $\widehat{Q}$ and $\widehat{\mathbf{R}}$ are self-adjoint
operators on $H$ that depend on the state $\psi ;$ $\mathbf{B}$ is a smooth
Markov process with components $B_{k}$, $k=1,2,3$ satisfying (here $\gamma $
is a new fundamental constant):%
\begin{equation*}
\overline{dB}_{i}(t)=0,\overline{\text{ }dB_{k}(t)dB_{j}(t)}=\delta
_{kj}\gamma dt.
\end{equation*}

\noindent The over-bar denotes averaging with respect to the underlying
probability space of the Markov process. Each $B_{k}$ is a map from its
index set (time) to random variables (measurable functions) on this space. A
probability distribution here as always - mathematically - is given by a
measure on a $\sigma -$algebra of sets (a Borel space). What this measure
corresponds to physically (Question (1)), and why these values of the
measure should dictate subjective probabilities (Question (2)), is as
obscure as ever.

Is there then no improvement in clarity about probability in this theory? It
would be odd if so; with a stochastic, indeterministic theory, the
probabilities are supposed to be in the theory at the ground level. One
would think that there be a clearer ontological basis to them than in a
deterministic theory. There is this difference: whereas in classical
statistical mechanics the measure is defined over a space of one-time data
(Cauchy data), each point of which encodes an entire trajectory, in a
stochastic theory the measure is defined over the space of trajectories
(histories). This, mathematically speaking, is the whole of the difference
between a deterministic and an indeterministic theory. The customary
distinction between epistemic\ and non-epistemic\ interpretations of
probability is more rooted in temporal matters than first appears. One may
say of a stochastic theory as much as of a deterministic one that
probabilities are epistemic: they reflect one's ignorance of which history
is ours. The difference \ is that in the deterministic case one must be
ignorant of much more, if probabilities are to make any sense: one cannot
know (in complete detail) even a single instant of our actual history.

What of the Born rule? It is built into the dynamics, Eq.(1), in the
dependence of the operators $\widehat{Q}$ and $\widehat{\mathbf{R}}$ on the
state. This is of the form:%
\begin{equation*}
\widehat{\mathbf{R}}(\psi )=\widehat{\mathbf{A}}-<\psi ,\widehat{\mathbf{A}}%
\psi >,\text{ }\widehat{Q}(\psi )=-\frac{i}{\hbar }\widehat{H}-\widehat{%
\mathbf{R}}\cdot \widehat{\mathbf{R}}
\end{equation*}%
where $\widehat{H}$ is the Hamiltonian, and $\widehat{\mathbf{A}}$ is a
commuting triple of self-adjoint operators transforming as a Galilean
3-vector (remember we only have a non-relativistic stochastic theory).
Different choices of \ the dependence of $\widehat{Q}$ and $\widehat{\mathbf{%
R}}$ on $\psi $ will lead to a stochastic process yielding statistics at
variance with the Born rule. It is therefore clearer in a state-reduction
theory than in pilot-wave theory why the squared amplitudes matter to
probabilities, since they enter directly into the dynamics. There is no
longer the possibility that the dynamics can reliably drive an initial state
through a sequence of states in which the relative frequencies of outcomes
is independent of the squared amplitudes (the Born rule). But everything now
hangs on the notion of a \textquotedblleft reliable\textquotedblright\
dynamics; the point is the same as before, that one might as well speak of a
\textquotedblleft typical\textquotedblright\ trajectory (and of probability
as a matter of ignorance as to which trajectory is one's own). And, indeed,
from a purely mathematical point of view, one is back to a definition of
probability as a measure on history space. What it is about a particular
history (the one that is actually realized) that makes it true that the
probability of a chance outcome has a particular value is never explained.
The link with subjective probabilities is as opaque as ever.

\section{Decoherence theory}

Returning to the conventional formalism, there are two important lessons
that we can draw from pilot-wave theory and state-reduction theory. In each
there is a universal state that applies to closed systems; and in each there
is state-reduction - merely \textquotedblleft effective\textquotedblright\
in the former theory, fundamental in the latter - yielding states which are
well-localized in configuration space (there is a \textit{single} preferred
basis). With this it is enough to recover all extant experimental data (at
least in the case of non-relativistic applications of quantum mechanics).
These are the lessons that are taken to heart in decoherence theory.

Decoherence theory was powered by investigations in open-quantum systems
theory just as the theory of quantum stochastic processes was. It has in
common with pilot-wave theory the assumption that the unitary equations of
motion are both fundamental and universal, and it has in common with
state-reduction theories many of the same equations, but derived as \textit{%
effective} equations, concerning only certain dynamical variables and
degrees of freedom of a system. In any given case of decoherence
probabilities are defined only for a certain basis or Boolean sublattice of
projectors. For systems at ordinary temperatures in which massive degrees of
freedom are weakly coupled to large numbers of much lighter ones the basis
for which simple, effective equations exist for the massive ones is always
(approximately) the same: it is given by projectors onto well-localized
regions of the configuration space of the massive degrees of freedom, at
least down to atomic dimensions and time-scales of the order of classical
thermal relaxation times. Thereby, so long as velocities and momenta are
obtained by averaging over these timescales, one obtains coarse-grained
trajectories in configuration space and phase space as well. These are
classically perspicuous descriptions of atomic and molecular processes (with
quantum correction terms added).

Decoherence plays a role in all the major schools of foundations in quantum
mechanics. It matters to the pilot-wave theory, for it explains just where
and when one can use the reduced state, discarding components of the state
(the \textquotedblleft empty waves\textquotedblright ) for the purposes of
actually applying the theory and computing trajectories. It matters to
state-reduction theory, which stochastically degrades every component of the
state save one. Degrade them too soon (before they have decohered) and the
predicted probabilities will differ from those of quantum \ mechanics. And
it matters to the consistent histories approach. Histories\textit{,} in this
formalism, are ordered sequences of projectors $\widehat{P}_{k}$ (of some
given resolution of the identity), interpreted as a sequence of events in
time, as specified by some discrete time variable and the resolution of the
identity used at each time. They are \textit{consistent} insofar as the
probabilities \ for each history are non-interfering, meaning the
probabilities are the same whether they are computed assuming the state is
reduced at each step (on sequential \textquotedblleft
measurement\textquotedblright\ of each projector $\widehat{P}_{k}$), or
assuming it is always the uncollapsed state. Whether or not a set of
histories is consistent depends on the basis used at each time, the state,
and on the unitary dynamics.

Components of the state decohere, in some basis, when there exist effective,
local equations of motion, propagating data along individual branches of the
state (referred to that basis), in approximate agreement with (and usually
adding small corrections to) the results of the conventional formalism using
the measurement postulates. It goes without saying that such histories are
approximately consistent - interference effects between histories could
hardly be significant if there are such effective in-branch equations. In
fact consistency is a much weaker requirement. One can define consistent
histories for the simplest imaginable systems, with only a handful of
degrees of freedom. One can represent the motion of a single electron in an
inhomogeneous magnetic field (the Stern-Gerlach experiment) in accordance
with various bases, each of them consistent; one can even smoothly modify
one basis into another whilst maintaining consistency \cite{Dowker and Kent}%
. Non-interference by itself is not enough to guarantee that any interesting
physics - effective equations - attaches to these histories. In the language
of Gell-Mann and Hartle \cite{Gell-Mann and Hartle}, decoherence proper
concerns a \textit{quasiclasical domain} and not just a consistent history
space.

Finally, decoherence matters to the Everett interpretation. For this
interpretation, like all the theories so far considered, there is a
wave-function of the universe, and like the consistent histories theory, the
unitary dynamics alone is fundamental. As in all of these theories there is
a preferred basis - defined by decoherence. We shall say more about the
Everett interpretation shortly.

Defining probability in terms of decoherence, one has more or less the same
Boolean sublattice of projectors throughout - a coarse-graining of
projectors on configuration space (equivalently, on phase space). But with
that it is clear that the premise of Gleason's theorem need not apply -
indeed, that it has no motivation whatsoever. If probability only makes
sense in the context of decoherence, which only arises for certain dynamical
variables and in certain situations, why suppose that probabilities can be
defined for\textit{\ }arbitrary\textit{\ }resolutions of the identity, with
a non-contextual additivity requirement built in from the beginning? Why
suppose probability has any meaning at all in regimes in which dynamical
decoherence does not exist? But as we shall see, the Born rule can be
derived from alternative premises, that from the point of view of
decoherence theory are very natural.

\section{A New Derivation of the Born Rule}

If probabilities only arise in the context of decoherence - if they are
\textquotedblleft emergent\textquotedblright\ - then they will have to have
some of the key attributes of decoherence.

\begin{itemize}
\item The first is that decoherence typically involves highly degenerate,
indeed infinitely degenerate, projectors (a projector onto any non-zero
volume of configuration space must have infinite-dimensional range). Call
this the \textit{degeneracy }condition.

\item The second is that decoherence is only \textit{approximate}; there is
no precise boundary below which the probabilities are undefined; except at
very low temperatures, there are many orders of magnitude over which
projectors can be further fine-grained without loss of decoherence. Call
this the \textit{stability }condition. \ 

\item The third is that the probability rule should be basis-independent; it
is the intrinsic relationship between the universal state and the Boolean
lattice of projectors that matters. Call this the \textit{invariance }%
condition.
\end{itemize}

Let us make this more precise, taking, for the sake of definiteness, for
configuration space $C,$ a family $F(C)$ of coarse-grainings $\Delta $ of $C$%
, with the natural partial ordering given by the inclusion relation on $C.$
For each $\Delta \in F(C)$ one has an associated Boolean sublattice $%
B_{\Delta }$ (generated by projectors onto the cells of $\Delta $) of the
lattice of projectors on the total Hilbert space $H.$ We are looking for a
probability measure $\mu :H\times B_{\Delta }\rightarrow \lbrack 0,1],$ $%
\Delta \in F(C)$ on projectors of infinite dimension which is intrinsic to $%
H $ and stable under variation of $\Delta ..$So we require:

(i) Each $\widehat{P}\in B_{\Delta },$ $\Delta \in F(C)$, $\widehat{P}$ has
infinite range (degeneracy)

(ii) For any unitary map $U:H\rightarrow H$, $\mu _{\Delta }(\psi ,\widehat{P%
})=\mu _{\Delta }(U\psi ,U\widehat{P}U^{-1})$ (invariance).

(iii) For any $\widehat{P}\in B_{\Delta }\cap B_{\Delta ^{\prime }},$ $%
\Delta ^{\prime },\Delta \in F(C)$, $\mu _{\Delta }(\psi ,\widehat{P})=\mu
_{\Delta ^{\prime }}(\psi ,\widehat{P})$ (stability).

\noindent To these we shall eventually add a continuity assumption. Our
claim is that under these assumptions, for sufficiently large families $F(C)$
of coarse-grainings of $C$, the Born rule follows uniquely. The proof in the
case of regular polyhedra at \textit{every }scale in $C$ (denote $F_{\infty
}(C)$) is particularly simple, although it has the disadvantage that it is
unphysical; at sufficiently small scales decoherence inevitably fails. We
shall later consider whether this idealization is really troublesome.

Certain results follow independent of any assumptionson $F(C)$. First a
definition. Call a set of orthonormal vectors $\{\varphi _{k}\}$ \textit{%
separating }for a set of disjoint projectors $\{\widehat{P}_{k}\},$ $%
k=1,...,d,$ if $\widehat{P}_{j}\varphi _{k}=\delta _{jk}\varphi _{k}$. Then:

\begin{lemma}
Let $\{\varphi _{k}\}$, $\{\widehat{P}_{k}\},$ $k=1,...,d$ , be separating, $%
\widehat{P}_{k}$ $\in B_{\Delta },$ and let $\psi
=\sum_{k=1}^{d}c_{k}\varphi _{k}$ . If $\mu $ satisfies (ii) and $\psi
^{\prime }=\sum_{k=1}^{d}|c_{k}|\varphi _{k}:$%
\begin{equation*}
\mu _{\Delta }(\psi ,\widehat{P}_{j})=\mu _{\Delta }(\psi ^{\prime },%
\widehat{P}_{j})\text{, }j=1,...,d.
\end{equation*}
\end{lemma}

\begin{proof}
Let $c_{k}=\exp (i\theta _{k})|c_{k}|$, $k=1,...,d$, and let $U_{\theta
}:\varphi _{k}\rightarrow \exp (-i\theta _{k})\varphi _{k}$, $U_{\theta }%
\widehat{P}_{k}U_{\theta }^{-1}=\widehat{P}_{k}$, $k=1,...,d$ (such a $%
U_{\theta }$ can always be constructed); the result is immediate from (ii).
\end{proof}

Likewise, the overall phase of the vector $\psi $ is irrelevant for the
probabilities. Now for the equiprobable case:

\begin{lemma}
As in Lemma 2, but let $|$ $c_{k}|^{2}=$constant. If $\mu $ satisfies (i),
(ii):
\end{lemma}

\begin{equation*}
\mu _{\Delta }(\psi ,\widehat{P})=\left\{ 
\begin{array}{c}
\frac{1}{d},\text{ }\widehat{P}\in \{\widehat{P}_{k}\},\text{ }k=1,...,d \\ 
\text{ \ \ }0\text{, }\widehat{P}\text{ orthogonal to }\sum_{k=1}^{d}%
\widehat{P}_{k}%
\end{array}%
\right. .
\end{equation*}

\begin{proof}
By Lemma 2, without loss of generality we may assume the $c_{k}$'s are all
real: $\psi =const.(\varphi _{1}+...+\varphi _{d}).$ First assume $d>1.$
Define $U_{\pi }$ by $U_{\pi }\varphi _{k}=\varphi _{\pi (k)}$, $U_{\pi }%
\widehat{P}_{k}U_{\pi }^{-1}=\widehat{P}_{\pi (k)}$, where $\pi $ is a
permutation (such a $U_{\pi }$ can always be constructed, since by (i) every
projector has the same dimension). Since $U_{\pi }\psi =\psi $, from
invariance: 
\begin{equation*}
\mu _{\Delta }(\psi ,\widehat{P}_{j})=\mu _{\Delta }(\psi ,\widehat{P}_{\pi
(j)}).
\end{equation*}%
Choose any $\widehat{P}_{k}$ (say $k=1$) and define $\widehat{P}_{1}^{\prime
}=\widehat{P}_{1}+(I-\sum_{j=1}^{d}\widehat{P}_{j}).$ Evidently $\{\varphi
_{j}\}$ is separating for $\widehat{P}_{1}^{\prime },\widehat{P}_{2},...,%
\widehat{P}_{d}$ and by the same argument $\mu _{\Delta }$ is constant on
this set as well. Since $\widehat{P}_{1}^{\prime }+\widehat{P}_{2}+...+%
\widehat{P}_{d}=I$ and $\mu _{\Delta }$ is a probability measure $\mu
_{\Delta }(\psi ,\widehat{P}_{j})=\frac{1}{d}$, $j=2,...,d$. The \ same
argument for any other choice of $k$ yields $\mu _{\Delta }(\psi ,\widehat{P}%
_{1})=\frac{1}{d}$; from additivity again, if $\widehat{P}$ is any projector
in $B_{\Delta }$ orthogonal to all the $\widehat{P}_{k}$'s then $\mu
_{\Delta }(\psi ,\widehat{P})=0.$ The case $d=1$ does strictly speaking
involve an additional (but very weak) assumption: that there are at least $3$
disjoint projectors in $B_{\Delta }$ disjoint from $\widehat{P}_{1}$, denote 
$\widehat{P}_{2},\widehat{P}_{3}$,$\widehat{P}_{4}$. Let $\widehat{P}%
_{4}^{\prime }=1-\sum_{k=1}^{4}\widehat{P}_{k}$. Since by assumption $%
\widehat{P}_{k}\psi =0$, $k=2,3,4$ $\mu (\widehat{P}_{2})=\mu (\widehat{P}%
_{3})=\mu (\widehat{P}_{4}^{\prime })$, by the result already proved (using
a permutation that leaves $\widehat{P}_{1}$ invariant). Let $\widehat{P}%
_{2}^{\prime }=\widehat{P}_{2}+\widehat{P}_{3}$; again $\widehat{P}%
_{2}^{\prime }\psi =$ $\widehat{P}_{4}^{\prime }\psi =0$ so $\widehat{P}%
_{2}^{\prime }$ and $\widehat{P}_{4}^{\prime }$ are equiprobable. But then $%
\widehat{P}_{2}^{\prime }$ and $\widehat{P}_{2}$ are equiprobable, and from
additivity $\mu (\widehat{P}_{2})=0$. By the same argument $\mu (\widehat{P}%
_{3})=\mu (\widehat{P}_{4}^{\prime })=0$, so by additivity $\mu (\widehat{P}%
_{1})=1.$
\end{proof}

The case $d=1$ is the \textit{eigenvector-eigenvalue rule}; this result and
the \ method of proof follows closely the mathematical ideas introduced by
Deutsch \cite{Deutsch} and Wallace\cite{Wallace}, \cite{Wallace3}. The next
two lemmas and Theorem 6 differ in certain respects, however. We shall come
back to the Deutsch-Wallace theorem shortly.

We need the stability condition to go beyond the equiprobable case. The
proof for $F(C)=F_{\infty }(C)$, where $C$ is $R^{n}$ and $H$ is isomorphic
to $L^{2}(R^{n},dx^{n}),$ is as follows. $($Note that condition (i),
degeneracy, is no longer needed as an independent assumption.) \ {}

\begin{lemma}
For any $\psi \in H=L^{2}(R^{n},dx^{n}),$ $\Delta \in F_{\infty }(C),$ and
for any $\widehat{P}\in B_{\Delta }$ and any integer $m$, there exists a
refinement $\Delta ^{\prime }$ of $\Delta $ and orthogonal projectors $%
\widehat{P}_{j}\in B_{\Delta ^{\prime }},$ $j=1,...,m$, summing to $\widehat{%
P}$, such that $\left\vert \widehat{P}_{j}\psi \right\vert $ is constant.
\end{lemma}

\begin{proof}
Let $\psi ^{\prime }=\widehat{P}\psi \neq 0$ (if zero, the proof is
immediate). For $n=1$, $\int_{-\infty }^{r}\overline{\psi }^{\prime }\psi
^{\prime }dx$ is a non-negative increasing function of $r$. By the
intermediate value theorem, there are real numbers $r_{1},...,r_{m-1}$ such
that $\int_{r_{j}}^{r_{j+1}}\overline{\psi }^{\prime }\psi ^{\prime }dx=const%
\frac{1}{m}$, $j=0,...,m-1$, $r_{0}=-\infty ,$ $r_{m}=\infty $. Choose as
projectors the characteristic functions $\chi _{\Delta _{j}^{\prime }}$ on $R
$, where $\Delta _{j}^{\prime }=[r_{j},r_{j+1}].$ The generalization to
higher dimensions is obvious.
\end{proof}

We may then prove

\begin{lemma}
Let $\mu $ be a probability measure on $B_{\Delta }\in F_{\infty }(C)$
satisfying (ii) and (iii)$.$ Let $\{\varphi _{k}\},$ $\{\widehat{P}_{j}\},$ $%
k=1,...,d$ be separating. Let $\psi =const\sum_{k=1}^{d}\sqrt{m_{k}}\varphi
_{k},$ $m_{k}\in Z$. Then:%
\begin{equation*}
\mu _{\Delta }(\psi ,\widehat{P})=\left\{ 
\begin{array}{c}
\frac{m_{j}}{\sum_{k=1}^{d}m_{k}},\text{ }\widehat{P}=\widehat{P}_{j},\text{ 
}j=1,...,d \\ 
0,\text{ }\widehat{P}\text{ orthogonal to }\sum_{k=1}^{d}\widehat{P}_{k}%
\end{array}%
\right. .
\end{equation*}
\end{lemma}

\begin{proof}
By Lemma 4, we may choose a fine-graining $\Delta ^{\prime }\in F_{\infty
}(C)$ of $\Delta $ such that for each $k=1,...,d$, $B_{\Delta ^{\prime }}$
contains $m_{k}$ orthogonal projectors $\widehat{P}_{k}^{j}$ summing to $%
\widehat{P}_{k}$, satisfying $\left\vert \widehat{P}_{k}^{j}\varphi
_{k}\right\vert =$ \textit{const}. Define $\varphi _{k}^{j}=\widehat{P}%
_{k}^{j}\varphi _{k}$, then $\varphi _{k}=\frac{1}{\sqrt{m_{j}}}%
\sum_{j=1}^{m_{k}}\varphi _{k}^{j}$ and $\psi
=const.\sum_{k=1}^{d}\sum_{j=1}^{m_{k}}\varphi _{k}^{j}$. By construction $%
\{\varphi _{k}^{j}\}$ is separating for $\{\widehat{P}_{k}^{j}\}$ ($%
m=\sum_{k=1}^{d}m_{k}$ in all) and the conditions of Lemma 3 apply; so $\mu
_{\Delta }(\psi ,\widehat{P}_{k}^{j})=\frac{1}{m}$ and $\mu _{\Delta }(\psi ,%
\widehat{P}_{k})=\mu _{\Delta }(\psi ,\sum_{j=1}^{m_{k}}\widehat{P}_{k}^{j})=%
\frac{m_{k}}{m}.$
\end{proof}

It is a short step to the general case. We need only assume that $\mu
_{\Delta }$ is continuous as a map $H\rightarrow \lbrack 0,1]$ (for fixed $%
\widehat{P}\in B_{\Delta }$). We thus obtain:

\begin{theorem}
$\bigskip $Let $\mu $ be as in Lemma 5 and for each $\widehat{P}\in
B_{\Delta }$ let $\mu _{\Delta }(.,\widehat{P}):H\rightarrow \lbrack 0,1]$
be continuous in norm. Then for any $\psi \in H$:
\end{theorem}

\begin{equation*}
\mu _{\Delta }(\psi ,\widehat{P})=\frac{<\psi ,\widehat{P}\psi >}{<\psi
,\psi >}.
\end{equation*}

\noindent The proof proceeds by constructing, for any $\psi $, a sequence of
vectors in $H$ of the form assumed in Lemma 5 that is separating for $%
\widehat{P}$ and $I-\widehat{P}$ that converges to $\psi $.

Having stated the theorem, two caveats. The first is that since it assumes
continuity in norm, its mathematical interest is considerably diminished.
One of the remarkable things about Gleason's theorem is that continuity of
the measure is \textit{derived}. But from a physical point of view, if
probabilities depend at all on vectors in $H,$ they surely vary continuously
with them. The assumption is physically perfectly natural.

The second is the one already noted, that $F_{\infty }(C)$ is unphysical.
But we should be clear why the idealization was needed. It is because, in
Lemma 5, the integers $m_{k}$ arising may be arbitrarily large, so the
number of orthogonal projectors required to sum to each $\widehat{P}_{k}$
must be arbitrarily large. That is only possible if we allow
coarse-grainings of $C$ that are arbitrarily small.

Suppose the scale of the coarse-graining is bounded below; what sort of
restriction does this place on these numbers? We are only interested in
exploring the probabilitistic structure of the state at the decoherence
lengthscale and above (for probability, if emergent, has no meaning at
smaller lengthscales). So we may suppose the state is approximately uniform
over some region $\Delta _{k}$ of configuration space, near the threshold of
decoherence, at the lengthscale $2l$; let $\Delta ^{\prime }$ \ be a
refinement of $\Delta $ at the lengthscale $l$; how many disjoint projectors
in $B_{\Delta ^{\prime }}$ are there, summing to the projector on $\Delta
_{k}$? The answer, for configuration space of dimension $n$, in the case of
hypercubes, is $2^{n}.$ So even taking the limits of decoherence into
account, we can derive rational ratios of probabilities using very large
integers - numbers that increase exponentially with the number of degrees of
freedom. Given our general philosophy, that probabilities are only defined
given decoherence and that they should be robust under changes of coarse
graining, we can legitimately demand that the distribution $\mu _{\Delta
}(\psi ,)$ should be smooth and not only continuous under variations in $%
\psi $, at least for macroscopic systems of large numbers of degrees of
freedom. It should be effectively constant over variations in ratios of
norms of one part in $2^{10^{22}}$.

We come back to the dependence of the Born rule on the \textit{purpose}\ of
the experiment. Although \ we do not yet have a clear picture of how to
interpret quantum mechanics using decoherence theory, we have an unambiguous
answer to this question. It is \textquotedblleft
behaviour\textquotedblright\ and not \textquotedblleft
purpose\textquotedblright ; it is a matter of what, at the sub-decoherence
level, as described in pure quantum mechanics, is reliably correlated (by
the measurement interaction) with decohering variables. It is only by virtue
of these correlations that probability as ermerging with decoherence has any
meaning at the microscopic level.

Here the details are familiar; they follow, in the simplest cases, the von
Neumann treatment of measurement processes. The measurement interaction
brings about correlations between projectors in $B_{\Delta }$, $\Delta \in
F(C)$, with projectors onto eigenspaces of dynamical variables of individual
subsystems $a$,$b,c,...$ . , described by (possibly finite--dimensional)
subspaces $H^{a}$ $\subset $ $H$. The only limit to this process is the
ingenuity of the experimenter. In the case of spin systems of small
dimension, it is a plausible claim that in this way one can experimentally
realize correlations between projectors in $B_{\Delta }$ and \textit{%
arbitrary} projectors on $H^{a}$.

So long as projectors in $B_{\Delta }$ and on $H^{a}$ can be reliably
correlated in this way - depending on the ingenuity of the experimenter -
probabilities assigned to projectors in $B_{\Delta }$ can be assigned to
projectors on $H^{a}$ as well. That is the whole story about probability at
the sub-decoherence level. That these correlations are non-contextual
follows automatically.

\section{The Everett Interpretation}

Probability, if only defined in the context of decoherence, must be given by
the Born rule. But what is the underlying physical picture? We have spoken
of quantum mechanical models of the experimental apparatus, applying quantum
mechanics directly to macroscopic systems, but of course decoherence theory
does not in itself solve the conceptual problems that follow from this. Lack
of clarity on this score makes it hard to answer the questions we are
concerned with: (1) What is objective probability? and (2) Why should
subjective expectations track these objective probabilities?

If we want clarity as to questions of what exists, we had better look to a
realist solution to the problem of measurement. If we want probability to
arise only in the context of decoherence, we had better not modify or add
new elements to the unitary formalism. This narrows down the available
alternatives. There are versions of the consistent histories interpretation
that may lay claim to a realist status, but those in which only a single
history is realized necessarily forsake the approximate character of
decoherence (essential to the derivation of the Born rule that we have
given), and require instead some new input to the theory in order to single
out a unique history space (to which the one and only history actually
realized belongs). The idea of environment super-selection rules and the
interpretation of an improper mixture (arrived at by tracing out
environmental degrees of freedom) in terms of ignorance has been dropped
even by its advocates \cite{Zureck2}, \cite{Zureck}.

That leaves only the literalist interpretation of the state, in which all
the branches are physically real. With that we are led to many worlds and to
the Everett interpretation: worlds are described by the components of the
universal state referred to the decoherence basis. As such, under the
unitary dynamics, the evolution from a single component of this basis into a
superposition is the evolution of one world into many. Worlds in this sense
divide.\footnote{%
They may also, in principle, recombine. It has long been recognized that
probability and the arrow of time are intimately related. This relation
leads on to others \cite{Zeh}; we cannot do justice to them here.} A chance
process is one in which a system is subject to division in this sense.

\subsection{Understanding branching}

Our objective here is not to evaluate solutions to the measurement problem,
only the status of probability within them. In Everett's approach, there is
now a clear cut answer to (1): probabilistic events arise only by branching.
Branching, or equivalently, the development of a superposition (referred to
the decoherence basis), is the basis of all objective physical indeterminism
(for quantum mechanics is taken to be both universal and fundamental). The
moment of branching is, to use Heisenberg's language, the point at which
\textquotedblleft potentiality\textquotedblright\ becomes \textquotedblleft
actuality\textquotedblright . Chances, as quantities, are squares of the
norms of the associated transition amplitudes - all categorical physical
properties and relations.

Just as important, branching (and therefore this transition) inherits the
approximate character of decoherence. One can put this in terms of \textit{%
vagueness} - that branching is vague, with clear-cut instances but no sharp
boundaries. Vagueness permeates ordinary language, but it is pervasive in
scientific theories as well. There is no precise physical definition of
tables or chairs, no more than of cells or molecules. Vagueness is endemic
in the chain of reduction, from ordinary objects to material science, the
solid state, and chemistry; from zoology and anatomy to molecular biology
and biochemistry. According to the Everett interpretation, extracting
quasiclasical phenomenology from the unitary dynamics of quantum mechanics
is subject to the same kinds of equivocation as confront any program for
recovering higher-level laws from more fundamental ones\cite{Wallace2}. The
methodological issues are all precisely the same.

If chances arise with branching, but branching depends on the details of the
coarse-graining, then chances can only be stable under variations in coarse
graining if they satisfy (iii), and hence (with assumptions (i), (ii)) the
Born rule - this the result just proved. It replaces Gleason's theorem;
probability is not assumed from the outset to be non-contextual and defined
for any basis; it is not assumed to have any fundamental significance at
all. Probability is \textquotedblleft emergent\textquotedblright .

One might object that the answer to Question (1) is then not so clear-cut
after all; chances arise with branching, but branching, because imprecisely
defined, is hardly being accounted for by any precise properties and
relations. But the same is true of paradigm cases of successful inter-theory
reduction. Reduction is never precise. It is not as though there should be
some precise and unique frequency distribution in electromagnetism that
corresponds to the colour \textquotedblleft red\textquotedblright , for
example. The point about the reduction in the case of chance is that it be
to \textit{categorical} properties and relations (that are not themselves
indeterminate, borderline, or chancy); it is that the substrate posited by
the reducing theory (the spectrum of waves\textit{)} should not employ
concepts just as mirky as those we sought to elucidate (the colour red%
\textit{)}. This is where Popper went wrong with his account of chance in
terms of dispositions.

In fact, in the special case of laboratory experiments (or more generally of
\textquotedblleft interpreted\textquotedblright\ phenomena), a more abstract
notion of branching is available that is reasonably precise; here one 
\textit{defines} branches, by sheer stipulation, in $1:1$ correspondence
with the different \textit{numbers }assigned to measurement outcomes (as
equivalence classes of configurations of the experimental display, that are
all taken to represent the \textit{same} numerical outcome). The number of
branches is the number of possible pointer positions on the dial. Of course
there still remain problems of borderline cases, if for no other reason than
that an experiment is always subject to inefficiencies and is always prone
to malfunction; but at this level, concerning branches that we count as
differing in clear-cut respects, we will be down to a small and finite
number. We will be perfectly able to make sense of their number.

The Everett interpretation does well with (1). It does much better than the
pilot-wave theory, even though the latter has all the resources of the
Everett interpretation \ - and then some, for it postulates additional
structure, namely a particular trajectory. But that is just where the
trouble comes in (when one introduces the trajectory); the trajectory may be
one in which the statistics are completely different from those predicted by
the Born rule. What does probability mean in such a case?

Some have thought that precisely the same worry arises in the Everett
interpretation. There too there exist \textquotedblleft anomalous
branches\textquotedblright , in which the recorded statistics do not match
the ones predicted by the Born rule. But there is an important difference.
According to Everett, there is nothing about a branch of this kind that can
ensure it will \textit{continue} to violate the Born rule (for there is no
fact of the matter as to what will happen following on from a given branch,
so long as every branch is given to division), unlike the situation for
anomalous trajectories in pilot-wave theory. Anomalous branches, in the
Everett interpretation, are like statistically anomalous segments, each of
finite length, in an infinitely extendible sequence of random numbers. They
have to exist if the sequence is to be genuinely random, but in no sense is
any given subsequence likely to \textit{continue} to be anomalous.

What about (2), the connection with subjective expectations? Why should the
amplitudes on branching be our guide for these? But at least it is clear
that on branching we \textit{ought }to be concerned with weights for
branches. For it is obvious that branching - \textit{personal} branching,
literally dividing in two, say - will lead to divided expectations, and this
will be so even given \textit{complete} knowledge of the branching process.
The two successors may differ widely, yet each will with as much right call
themselves the same person as before. In the face of branching there is no $%
1:1$ criterion of identity in the forward direction of time. But if one is
to make provision for one's successors, one must allocate resources among
them. And one can hardly do this without introducing weightings, implicit or
explicit, in one's reasoning. One cannot ready oneself for anything and
everything.

Philosophers have long disagreed on how, in the presence of branching,
questions of personal identity are to be settled \cite{Rorty}; we should
make no pretence that in this matter there is any real consensus. But the
one response that is really damaging to the Everett interpretation has found
few advocates: it is that in the face of branching one should expect \textit{%
nothing}, \textit{oblivion}. This view is inherently implausible, given that
each of my successors is \textit{ex hypothesi }functionally exactly the same
as me. Every successor has all of my attributes and memories; every
successor professes himself to be me on the basis of physical continuity
(and on every other physical criterion). No wonder this view has found few
supporters. 

\subsection{Deutsch's argument}

If not oblivion, then divided expectations. If divided expectations, then
divided how, and with what weighting? What preferences ought one to have for
one process of branching (for performing one choice of experiment), with a
given utility in each branch (for each experimental outcome), over another?
But just at this point Deutsch's argument comes into play.

Deutsch's strategy, following de Finnetti \cite{de Finetti}, and before him
von Neumann and Morgenstern \cite{von Neumann}, was to define subjective
probabilities (hereafter, \textit{weights}) in terms of the preferences of a
rational agent among a set of games $g\in G$ each with some set of outcomes $%
E_{k}$, $k=1,...,d$ with associated utilities (\textquotedblleft
payoffs\textquotedblright ) - concrete rewards, cash prizes say, that the
agent values - belonging to some set $\mathcal{U}$. Call $\mathcal{P}%
:E_{k}\rightarrow \mathcal{U}$ the \textit{payoff function} for that agent.
If rational, the ordering $\preceq $ on $G$ defined by one's preferences
should satisfy certain obvious rules (for example, transitivity). The
strategy is then to find a strong enough but still plausible set of rules
sufficient to ensure that for each game $g$ there exists real numbers $%
p_{k}\in \lbrack 0,1]$ for each outcome $\mathcal{\lambda }_{k}$, $%
k=1,...,d, $ summing to one, and quantities $\mathcal{V}(g)=%
\sum_{k=1}^{d}p_{k}\mathcal{P}(\mathcal{E}_{k}),$ such that $g\preceq
g^{\prime }$ if and only if $\mathcal{V(}g)\leq \mathcal{V(}g^{\prime })$.
If $G$ is big enough, indeed, one would hope to show that the numbers $p_{k}$
(weights) for the outcomes in each $g$ are unique. The important point in
this is that the $p_{k}$'s arrived at in this way will be \textit{%
independent }of an agent's utilities. A rational agent will act as if
attempting to maximize the expectation value of her utilities, using these
weights as probabilities. It is because of this representation theorem that
subjective probability is in so much better shape than objective
probability. If this is what probability is, one can explain why it obeys
the rules that it does.

Deutsch's remarkable claim is now that the preference ordering of rational
agents, in the face of quantum games, can be so constrained that the weights
defined by these preferences (independent of their actual utilities) agree
with the Born rule.

This result is so surprising that one wants to have an inkling of how it was
obtained. Here we shall follow Wallace \cite{Wallace}, who has substantially
revised and simplified the argument. A quantum game can be played using any
quantum experiment, simply by agreeing on various payoffs (positive or
negative) on each possible outcome. What is an experiment, according to
Everett? It is a special kind of process, involving stable macroscopic
objects, described by effective equations, such that states can be
attributed to sub-systems (typically molecular), as relative states, that
involve unitarily (i.e. as a product state) with respect to the apparatus
(this the state-preparation process). Following some unitary evolution
preserving this product structure, they evolve into an entanglement with the
measurement device. Components of this eventually include macroscopic
degrees of freedom (pointer positions).

From a mathematical point of view one introduces a tensor-product in the
Hilbert space for a particular branch, distinguishing some microscopic
sub-system $a$ with Hilbert space $H^{a}$ from all the rest. Suppose (for
simplicity) that $H^{a}$ has finite dimensions. The state preparation device
produces, in a reliable way, vectors in a certain subspace of $H^{a}$ (for
simplicity suppose 1-dimensional, so a particular state $\phi $), in a
tensor product with the state of the rest of the apparatus and its
environment. The entanglement subsequently brought about is with some set of
orthogonal states $\phi _{k}$ $\in H^{a},$ $k=1,...,d$ of $a,$ with
decohering states of the apparatus and environment (grouped together when
they give rise to the same pointer-reading). The \ latter reliably leave
behind them a macroscopic trace.

In these models it is useful to introduce numbers $\lambda _{k}$ for the
states in $H^{a}$ which have some dynamical significance - eigenvalues $%
\lambda _{k},$\ associated with eigenstates of some self-adjoint operator $%
\widehat{X}=\sum_{k=1}^{d}\lambda _{k}\widehat{P}_{\phi _{k}}$; these
replace the $E_{k}$'s above. The instrument display, meanwhile, registers
numbers concretely, so one has some definite assignment of the $\lambda _{k}$%
's with these numerals (usually taken as the identity). The experiment is
converted to a game by specifying a map from these numerals to an agent's
utilities in $\mathcal{U}$, which we can model directly in terms of the
pay-off function as a map $\mathcal{P}:Sp(\widehat{X})\rightarrow \mathcal{U}%
.$ Suppressing explicit reference to $H^{a}$, a quantum game is then given
by an ordered triple $<\phi ,\widehat{X},\mathcal{P}>.$

But adopting this schema, we must recognize the arbitrary elements in it. It
is obviously possible to compensate for a change in labels $\lambda _{k}$ by
a change in the pay-off function. This corresponds to a certain
arbitrariness in the choice of self-adjoint operator that the experiment is
said to measure: whether it is is $\widehat{X}$ with payoff function $%
\mathcal{P}$, or $f(\widehat{X})$ with payoff function $\mathcal{P\circ }%
f^{-1}$ (for some invertible $f$ on $Sp(\widehat{X}))$. For another
arbitrary element, typically the initial product state involving $\phi $
will be subject to a unitary evolution $U$ on $H^{a}$ which preserves the
product structure. Indeed, the preparation device may best be modelled using
a sequence of Hilbert spaces with intertwining operators $U:H^{a}\rightarrow
H^{b}$. From an Everettian standpoint it is now entirely arbitrary which of
these is taken to be \textit{the}\ initial state; the experiment can with as
much right be called a measurement of $U\widehat{X}U^{-1}$ in the state $%
U\phi $ as of $\widehat{X}$ in the state $\phi $. There is nothing in the
physics to say. It is only the correlations between vectors in $H^{a}$ with
payoffs or pointer readings in $\mathcal{U}$, that is relevant to an
experiment; not the vector at which time - this is entirely arbitrary.

We have established two principles. Under our schema for quantum games, the
triples $<\phi ,\widehat{X},\mathcal{P}>$, many games can be realized by a
single physical process. Since preferences among games should concern the
physical world rather than the models used to describe it, they should value
games as the same if they can be realized by the same physical process. Let $%
g\thicksim g^{\prime }$ if and only if $g\preceq g^{\prime }$ and $g^{\prime
}\preceq g$. We require for any unitary on $H^{a}$ and any invertible $f$ on 
$Sp(\widehat{X})$ the equivalence principles:

\begin{eqnarray*}
\mathbf{Payoff}\text{ }\mathbf{Equivalence} &\text{:}&<\phi ,\widehat{X},%
\mathcal{P}>\sim <\phi ,f(\widehat{X}),\mathcal{P\circ }f^{-1}> \\
\mathbf{Measurement}\text{ }\mathbf{Equivalence} &\text{:}&<\phi ,\widehat{X}%
,\mathcal{P}>\sim <U\phi ,U\widehat{X}U^{-1},\mathcal{P}>
\end{eqnarray*}

This is a schema well-suited to the Everett interpretation, but it can be
motivated on other grounds; it can even be motivated operationally \cite%
{Saunders}. Deutsch's decision theoretic axioms naturally make no reference
to the Everett interpretation. The analysis that follows can, therefore, be
largely freed from its dependence on Everett. But as we shall see, it then
fails to have the foundational significance for probability that we are
after. We shall come back to this point in due course.

First Deutsch's decision theory axioms. We will prove only one of his
results, and for that we only need two axioms. For simplicity, we assume
that $\mathcal{P}$ is linear (so $\mathcal{P(}x_{1}+x_{2})=\mathcal{P(}%
x_{1})+\mathcal{P(}x_{2})$ - this is a convention on the labels $\lambda _{k}
$). Let $f_{s}:R\rightarrow R$ be the function $f_{s}(x)=x+s,$ and let $%
-I:R\rightarrow R$ be $-I(x)=-x.$ The first axiom is:%
\begin{equation*}
\text{\textbf{Sure-thing principle}}:\mathit{Let\ }g=<\phi ,\widehat{X},%
\mathcal{P}>\mathit{,\ }g^{\prime }=<\phi ,\widehat{X},\mathcal{P}\circ
f_{s}>\mathit{;\ then\ }\mathcal{V(}g^{\prime })=\mathcal{V}(g)+\mathcal{P}%
(s).\mathit{\ }
\end{equation*}%
I am indifferent between receiving $\mathcal{P}(s),$ and then playing game $g
$, and playing $g$ and then receiving $\mathcal{P}(s),$ whatever the
outcome. But the latter is $g^{\prime }$.

The second axiom is:%
\begin{equation*}
\text{\textbf{Zero-sum\ rule}}:\text{\textit{Let} }g=<\phi ,\widehat{X},%
\mathcal{P}>,g^{\prime }=<\phi ,\widehat{X},\mathcal{P}\circ -I>;\text{%
\textit{then }}\mathcal{V(}g)=-\mathcal{V}(g^{\prime }).
\end{equation*}

\noindent It must be possible for I and my banker to share exactly the same
preferences, and to play the same game: what I am prepared to pay to play $%
g, $ I pay to him. The most I am prepared to pay should be the least he is
prepared to accept. But whereas I play $g,$ he plays $g^{\prime }.$

The rational for \ these principles can also be stated in a way that takes
branching explcitly into account. For the first, if I accept $\mathcal{P}(s)$
before playing $g$, each of my successors inherits $\mathcal{P}(s)$ as well
(for the utility too is subject to branching), and the situation at the end
is the same as if I had played $g^{\prime }$. For the second, if I am
prepared to swap \ with my banker before playing $g$, being paid what I
would otherwise have paid, each of my successors is swapped with my banker's
successors, and pays what he would otherwise have been paid; but the latter
is just $g^{\prime }.$

With that it follows that for $\widehat{X}=x_{1}\widehat{P}_{\phi _{1}}+x_{2}%
\widehat{P}_{\phi _{2}},$ $\mathcal{V(\phi }_{1}+\phi _{2},\widehat{X},%
\mathcal{P})=%
{\frac12}%
(\mathcal{P}(x_{1})+\mathcal{P}(x_{2}))$, in accordance with the Born rule.
For $\phi _{1}+\phi _{2}$ is invariant under the permutation $\pi $ of $\phi
_{1}$ with $\phi _{2}$, so by payoff equivalence $\mathcal{V(\phi }_{1}+\phi
_{2},\widehat{X},\mathcal{P})=\mathcal{V(\phi }_{1}+\phi _{2},U_{\pi }%
\widehat{X}U_{\pi }^{-1},\mathcal{P}).$ By measurement equivalence this is $%
\mathcal{V(\phi }_{1}+\phi _{2},\widehat{X},\mathcal{P\circ \pi }^{-1})$.
Since $\pi ^{-1}=\pi =$ $-I\circ f_{-x_{1}-x_{2}},$ by the sure-thing
principle and the linearity of $\mathcal{P}$ one obtains $\mathcal{V(\phi }%
_{1}+\phi _{2},\widehat{X},\mathcal{P\circ }-I)+\mathcal{P}(-x_{1}-x_{2})$.
By the zero-sum rule and linearity again this is $\mathcal{V(\phi }_{1}+\phi
_{2},\widehat{X},\mathcal{P\circ }-I)=-\mathcal{V(\phi }_{1}+\phi _{2},%
\widehat{X},\mathcal{P}).$ So $\mathcal{V}(\phi _{1}+\phi _{2},\widehat{X},%
\mathcal{P})=-\mathcal{V(\phi }_{1}+\phi _{2},\widehat{X},\mathcal{P})+%
\mathcal{P}(x_{1})+\mathcal{P}(x_{2}).$

Deutsch called this his \textquotedblleft pivotal result\textquotedblright ,
and for good reason: it is the first time that any rational basis has been
found to tailor one's subjective probabilities to the quantum mechanical
ones. It is also the first step towards proving a general principle. Observe
that the argument goes through for any $\phi \in H$ of the form $\phi
_{1}+\phi _{2}+c\phi _{3}$, where $\widehat{P}_{1}\phi _{3}=\widehat{P}%
_{2}\phi _{3}=0$; observe further that the antecedent can be stated as the
condition that the Born rule for $x_{1}\widehat{P}_{\phi _{1}}$ yields the
same value as for $x_{2}\widehat{P}_{\phi _{2}}$. So we have proved for any
orthogonal projectors:%
\begin{equation}
\text{\textbf{Special Equivalence: }If }\mu (\phi ,\widehat{P}_{1})=\mu
(\phi ,\widehat{P}_{2})\text{ then }<\phi ,\widehat{P}_{1},\mathcal{P}>\sim
<\phi ,\widehat{P}_{2},\mathcal{P}>.
\end{equation}%
Further axioms of decision theory are required to derive the analogous
condition in which the vectors in $H^{a}$ are different but the $\widehat{P}$%
's are the same. Combining the two, we have \cite{Wallace3}:%
\begin{equation*}
\mathbf{General}\text{ }\mathbf{Equvalence}:\text{If }\mu (\phi _{1},%
\widehat{P}_{1})=\mu (\phi _{2},\widehat{P}_{2})\text{ then }<\phi _{1},%
\widehat{P}_{1},\mathcal{P}>\sim <\phi _{2},\widehat{P}_{2},\mathcal{P}>.
\end{equation*}

Given the general equivalence condition, the full representation theorem
follows from decision-theoretic axioms that are exceedingly weak - axioms
which, in point of fact, should be acceptable whatever one's views on what
it is proper to believe in the face of personal division \cite{Wallace3}.
This full representation theorem is then none other than the principal
principle (as is the general equivalence rule in the case of
equiprobability). But there is an important difference, that the principle
has been derived even under the condition that the agent has perfect
knowledge. So it holds \textit{unrestrictedly}; there is no room for the
rider to the principle that a rational agent should be indifferent between
playing two games if the objective probabilities for the same utilities are
the same, whatever additional information she has, \textit{provided it does
not bear on the actual outcomes of the games} (a rider that has in fact
proved notoriously difficult to state with any great precision). There is no
need to exclude information of this sort because, of course, she knows
everything there is to know about the outcomes of the games. As Wallace has
stressed \cite{Wallace3}, an unrestricted principal principle cannot
possibly be accepted, let alone deduced, on any interpretation of quantum
mechanics in which only a single history is real; for why be indifferent
between two games, even if they have the same probability as given by the
Born rule, if you know their actual outcome as well; won't it depend on what
those outcomes actually are?

\subsection{Measurement neutrality}

The general equivalence condition can in fact be stated in a way that is
independent of any particular reference to experiments and independent of
the schema we have used. As Wallace has shown \cite{Wallace3}, the
derivation of the full representation theorem is all the simpler. Given
general equivalence, the representation theorem is altogether unproblematic;
the decision theory axioms are so weak that there is no need to consider
what it is proper to believe on personal division. And it may be, on the
basis of the Everett interpretation, that one can argue for the general
equivalence condition directly.

But there is something to be said for the argument we have just sketched. We
should hold fast to the belief that on fission, we should not anticipate
oblivion; that, as argued by Parfit, psychological continuity is what
matters, and not a relation with the formal properties of identity \cite%
{Parfit}. We should look forward to the same sort of first-person
perspective, whatever it is, that we do in the absence of branching. But
then our situation is one of subjective uncertainty, in the \ following
sense: there is no \textit{one} perspective that we should look forward to;
we must, in some sense, entertain them all, and we must make provision for
them all (or as many as we can reasonably survey), weighting them
appropriately - \textit{just as we do in the face of uncertainty}. It is
only in special situations - for example, when \textit{no} successor has
some outcome - that one is entitiled to ignore that outcome completely, and
give it zero weight.

These questions lead off into philosophy. The other sort of criticism that
can be made of the Deutsch-Wallace derivation of the general equivalence
condition (and the special equivalence condition as above) concerns the use
of the schema for quantum games. Is this schematization (subject to pay-off
and measurement equivalence) sufficiently detailed? Does it tell you
everything you really want to know in the context of decisions about
real-life experiments?

According to Everett, we certainly can characterize quantum games by triples
of the form $<\phi ,\widehat{X},\mathcal{P}>$ (although that is common
ground to a wide variety of approaches to foundations). Likewise \ the
Everett interpretation licences payoff equivalence and measurement
equivalence (although as mentioned, so can other approaches). But why should
a rational agent take absolutely nothing else into account in determining
her preferences? Mightened there be other features \ of quantum measurements
(or quantum games) that are worth taking into account?

One can simply \textit{deny} this possibility. A principle to this effectt
has been called \textit{measurement neutrality \cite{Wallace}}; it is the
suspicion that measurement neutrality is too strong, or rests on unwarranted
or incoherent assumptions as to what it is right to believe in the face of
personal division, that prompts one to seek a more \ direct argument for the
general equivalence condition. But better is to seek a direct argument for
measurement neutrality.

Why believe in this principle? What more can be said of a measurement
process, according to the Everett interpretation, not captured in the
schema, and what is the rational for ignoring it? There is of course a vast
amount of information that is\textit{\ not }contained in the schema. There
is \textit{everything else} happening in each branch; there is the world
outside the laboratory (assuming games are played in laboratories), and
there are all the detailed goings-on in the laboratory that were unmentioned
in the pay-off (including molecular goings-on that are not in fact
dectected). But if these are deemed to matter to her, let them be put into
her utilities, and let her prize those games whose pay-off functions include
them explicitly; still, if she is rational, her preferences will be
consistent with the Born rule, along the lines of the argument \ just given.
The real issue, it is now becoming clear, is that the one thing that she 
\textit{cannot} put into her pay-off function is the amplitude; it is not a
possible pay-off for her that can be arranged. For there is no dynamical
process, according to Everett, whereby the amplitude of a branch can be
measured, and its value exhibited by a display, or otherwise encoded into
the pay-off function $\mathcal{P}$. \textit{No} unitary evolution can ever
achieve that.

This goes a long way to explaining why the amplitudes should play the part
in our mental lives that they do, when it comes to our preferences
(appearing, distinctively, in the weights with which we view outcomes,
rather than as part of the outcomes), and it makes clearer the generality of
our schema (that it can in principle be applied to arbitrarily complex
initial conditions and pay-off functions). But \ it will not do to explain
why nothing else but amplitudes can matter to these weights. It is one thing
to care about details of the apparatus as goes the physical outcomes
produced by it \ - these should simply be entered into one's utilities. But
what about details of the apparatus that effect its dynamical functioning,
that are not captured in our simple schema?

There are of course a vast range of physical considerations bearing on the
detailed dynamics of an experiment, but it is reasonable to distinguish
between those that make a difference to the branching structure and those
that do not. We are speaking of a rational agent who may know everything
there is to know: for her, subjective probability, weights, arise in the
first instance on personal division.

Any dynamics not involving branching concerns the purely deterministic
development of the branch (insofar as branches can ever evolve
deterministically). Experiments modelled in the same way by our schema that
differ on this are likely to have different efficiencies and will differ in
their systematic errors and the ways they are prone to malfunction. There is
no reason why rational agents should regard them as precisely equivalent,
but equally, just because they concern deterministic effective processes,
they seem unrelated to the foundational questions about probability.

It is otherwise with distinctions among experiments that turn on differences
in branching. Again, they break down into two kinds. The first concerns the
precise details whereby an initial entanglement of microscopic system with
the measuring apparatus is produced, including the branching that takes
place as progressively larger numbers of degrees of freedom are entangled
with it. This takes place over decoherence timescales and is extremely
rapid. The second kind concerns subsequent branching unrelated to the
coupling of the apparatus to the microscopic system, usually considered in
terms of statistical fluctuations (and that go on continuously whether or
not any measurement are performed). Of course experiments attempt to control
for the latter - this is noise in the signal that is better or worse
eliminated - \ but in no sense is it possible to model this kind of
branching explicitly. It is dealt with again at the level of effective
equations.

We are left with the key arena in which the initial entanglement is
established and subsequently propagated to include large numbers of degrees
of freedom - the business of measurement theory proper. Is it not reasonable
that a rational agent treat differently experiments that differ in these
respects, even if they are modelled in the same way in our schema of quantum
games? This is the key objection to the derivation of the general
equivalence principle from decision theoretic principles: the amplitudes
(and weights as given by the Born rule) may be the same, but the branching
structure introduced by the measurement may differ wildly.

\subsection{Decoherence, again}

The objection that branch number may have a role in dictating preferences
has to be understood in the right way. It is not that the number of outcomes
counted as distinct when it comes to the pay-off function - equivalently,
the number of gradations on the instrument read-out - may not matter to our
preferences: that number is already explicit in our schema (in the
specification of the triples $<\phi ,\widehat{X},\mathcal{P}>$ ). It is
hard-wired in the instrument display. As such it is available to a rational
agent, to be incorporated in her utilities if she will. So she may favour\
quantum games with five outcomes rather than four, because five is her
favourite number; or she may dislike outcomes with thirteen in the display.
Anything physically realized in any branch can always be entered into her
utilities, and be looked after at the level of her payoff function, without
compromising her conduct as a rational agent - and therefore in accordance
with the Born rule. She will act irrationally, however, if she believes her
liking for fives is a reason to \textit{weight} outcomes of games with five
in the display as greater than those without, or to set the \textit{weight }%
of any outcome with thirteen in it to zero. If she does this, she will have
to violate the sure-thing principle or the zero-sum rule in some cases; or
to hold the consequences of the Everett interpretation for pay-off
equivalence and measurement equivalence to be false. Exactly the same
applies if she assumes that each outcome, corresponding to each gradation of
the display, should have equal weight: she will be convicted of
irrationality or ignorance of quantum mechanics.

The objection is not that. It is that the schema for measurements is leaving
something out of account, not that what is does contain can be acted on
irrationally. It is that for each outcome with each given pay-off, \textit{%
the number of branches all with that outcome} has been ignored. Like
amplitudes, and unlike the number of gradations on the instrument display,
this is not \ information that can be factored into one's utilities; these
are not numbers that can be reliably realized in a branch by any unitary
dynamics. All the more reason, then, to think that they may be relevant to
agent preferences in the way that amplitudes are, so relevant to her weights.

But the answer to this should now be obvious. \textit{There is no such thing
as this number}. The only significance it has in concrete physical terms is
what is coded up in the number of instrument gradations. it is true that one
can specify such a number theoretically, for a given choice of decoherence
basis, but it has no categorical physical significance; it is not part of
what is really there.

The reason no other number can be defined has already been rehearsed; it is
because decoherence is an imprecise concept. Formally, the number of
decohering branches corresponds to the number of decohering projectors -
and, one has to add, \textquotedblleft for a particular choice of coarse
graining on configuration space\textquotedblright . There is no such thing
as \textit{the finest decohering set of projectors}. The picture, in Everett
theory, of the wave function of the universe as an endlessly branching tree,
breaks up as one goes into the fine detail. It is no criticism of the
Deutsch-Wallace argument that it leaves out of account a physically
meaningless quantity. The same applies to the branching that takes place
constantly, independent of the measurement of the microscopic system \textit{%
per se} (\textquotedblleft background noise\textquotedblright ); there is no
such thing as the number of branches produced in this way either. It is no
part of any principle of rationality to take note of what is not there.

We have come full circle. A rational agent, who knows everything there is to
know about the physical world, will still have preferences among quantum
games, and she ought to order her preferences consistently. In so doing (for
a sufficiently rich set of games) she will act as if assigning a unique set
of weights to outcomes (independent of the utilities that she assigns to
them) that have to obey the rules of probability theory. If she believes
quantum mechanics to be true, under the Everett interpretation, she will
consider the schema for quantum games a reasonable idealisation of what goes
on in measurements, subject to outcome equivalence and measurement
equivalence. Moreover, she ought to believe in the sure-thing principle and
the zero-sum rule (although here there are weaker principles that will do as
well), so she ought to conclude from the equality of the norms of amplitudes
for outcomes to equality of the weights that she gives to them. She ought to
believe in the special equivalence condition. And so on, to the general
condition and the full representation theorem. But the reason she should
consider the schema for experiments adequate (although she ought to have
quibbles over the neglect of detector inefficiencies and the presence of
background noise) is because her subjective weightings depend on branching,
and branching depends on decoherence; it is because of what the chances are,
in physical terms, that there is no fact of the matter as to the number of
branches. And, conversely, it was exactly because decoherence is a matter of
approximation, that if chance is to emerge with decoherence, then it had
better be stable under changes in coarse-graining -\ equivalently, under
change in branch number - that we were able to isolate the ratios in modulus
squares of the amplitudes (for decohering projectors) as the only invariant
quantities that could play the role of chance.

Subjective and objective probability emerge at the end of the day as
seamlessly interjoined: nothing like this was ever achieved in classical
physics. Philosophically it is unprecedented; it will be of interest to
philosophers even if quantum mechanics turns out to be false, and the
Everett interpretation consigned to physical irrelevance; for the
philosophical difficulty with probability has always been to find \textit{any%
} conception of what chances are, in physical terms, that makes sense of the
role that they play in our rational lives.

Still, the Everett interpretation is inherently fantastic; one would like if
possible to free the argument from any dependence on it. Yet we encountered
again and again points on which the Everett interpretation played a critical
role - where the very features of the approach that make it unbelievable
were of special salience. Introduce additional elements, over and above the
unitary dynamics - whether hidden-variables or an additional stochastic
dynamics controlling the state - and the symmetries used to drive the proofs
for the individual case \textit{have} to be broken. Try to reformulate the
derivation so as to apply to probability distributions over ensembles, and
we are back to the same foundational questions about the latter as in
classical theory. And meanwhile the very conclusions of the argument become
obviously untenable: the unrestricted special equivalence condition that we
derived is incoherent if there is only a single history. The arguments we
have considered give no hope at all that one can derive the principal
principle on any basis but Everett's.

It is ironic that the interpretation of probability in the Everett
interpretation has always been thought to be its weakest link. On the
contrary, it seems that it is one of the strongest points in its favour.

\bigskip

\textbf{Acknowledgements }My thanks to Harvey Brown, Antony Valentini, and
William Demopoulos for stimulus and helpful suggestions. I owe a special
debt to David Wallace, whose work I have drawn on heavily, and to Michael
Dickson, to whom I owe the suggestion that a mathematical core of the
Deutsch-Wallace argument can be freed from any mention of experiments.

\end{document}